\pdfoutput=1
\documentclass[12pt,a4paper]{article}

\usepackage{ifthen} 
\newboolean{pdflatex}
\setboolean{pdflatex}{true} 

\newboolean{articletitles}
\setboolean{articletitles}{true} 

\newboolean{uprightparticles}
\setboolean{uprightparticles}{false} 

\newcommand{\RNum}[1]{\uppercase\expandafter{\romannumeral #1\relax}}

\def\splot{\ensuremath{s\mathcal{P}lot}\xspace}

\def\Sigmacpp {{\ensuremath{\Sigmares^{++}_\cquark}}\xspace}
\def\Sigmacstarpp {{\ensuremath{\Sigmares^{*++}_\cquark}}\xspace}
\def\DorDstarm {{\ensuremath{D{}^{(*)-}}}\xspace}
\def\DorDstarb  {{\ensuremath{\Dbar{}^{(*)}}}\xspace}
\def\DorDstarbz  {{\ensuremath{\Dbar{}^{(*)0}}}\xspace}

\def\scstpp {\ensuremath{\PSigma_c^{(*)++}}\xspace}

\def\LbtoLcDzK {\decay{\Lb}{\Lc \Dzb \Km}}
\def\Lbdecays {\decay{\Lb}{\scstpp D{}^{(*)-} \Km}}
\def\LbtoScDmK {\decay{\Lb}{\Sigmacpp \Dm \Km}}
\def\LbtoScDstK {\decay{\Lb}{\Sigmacpp \Dstarm \Km}}

\def\LbtoScstDmK {\decay{\Lb}{\Sigmacstarpp \Dm \Km}}
\def\LbtoScstDstK {\decay{\Lb}{\Sigmacstarpp \Dstarm \Km}}

\def\LbtoSDstK {\decay{\Lb}{\scstpp \Dstarm \Km}}
\def\LbtoSDmK {\decay{\Lb}{\scstpp \Dm \Km}}

\def\DzbtoKpi{\decay{\Dzb}{\Kp\pim}}
\def\DmtoKpipi{\decay{\Dm}{\Kp\pim\pim}}
\def\DstmtoDzbpi{\decay{\Dstarm}{\Dzb\pim}}

\newcommand{\RBRresult}[4]{\ensuremath{#1\pm#2\pm#3\pm#4}}
\newcommand{\RBRresultNoBrErr}[3]{\ensuremath{#1\pm#2\pm#3}}

\def\paperauthors{LHCb collaboration} 
\def\paperasciititle{First observation of Lambda_b0 -> Sigma_c(*)++ D(*)- K- decays}
\def\papertitle{First observation of \mbox{\Lbdecays} decays}
\def\paperkeywords{{High Energy Physics}, {LHCb}} 
\def\papercopyright{\the\year\ CERN for the benefit of the LHCb collaboration} 
\def\paperlicence{CC BY 4.0 licence}
\def\paperlicenceurl{https://creativecommons.org/licenses/by/4.0/}

\usepackage{tikzscale}
\usepackage{tikz-feynman}
\usepackage{setspace}

\usepackage[top=1in, bottom=1.25in, left=1in, right=1in]{geometry}

\usepackage{microtype}
\usepackage{lineno}  
\usepackage{xspace} 
\usepackage{caption} 

\usepackage{graphicx}  
\usepackage{color}
\usepackage{colortbl}
\graphicspath{{./}} 

\usepackage{amsmath} 
\usepackage{amssymb}
\usepackage{amsfonts}
\usepackage{upgreek} 
\usepackage{subcaption}

\newcommand*\patchAmsMathEnvironmentForLineno[1]{%
\expandafter\let\csname old#1\expandafter\endcsname\csname #1\endcsname
\expandafter\let\csname oldend#1\expandafter\endcsname\csname
end#1\endcsname
 \renewenvironment{#1}%
   {\linenomath\csname old#1\endcsname}%
   {\csname oldend#1\endcsname\endlinenomath}%
}
\newcommand*\patchBothAmsMathEnvironmentsForLineno[1]{%
  \patchAmsMathEnvironmentForLineno{#1}%
  \patchAmsMathEnvironmentForLineno{#1*}%
}
\AtBeginDocument{%
\patchBothAmsMathEnvironmentsForLineno{equation}%
\patchBothAmsMathEnvironmentsForLineno{align}%
\patchBothAmsMathEnvironmentsForLineno{flalign}%
\patchBothAmsMathEnvironmentsForLineno{alignat}%
\patchBothAmsMathEnvironmentsForLineno{gather}%
\patchBothAmsMathEnvironmentsForLineno{multline}%
\patchBothAmsMathEnvironmentsForLineno{eqnarray}%
}

\usepackage{hyperxmp}

\usepackage[pdftex,
            pdfauthor={\paperauthors},
            pdftitle={\paperasciititle},
            pdfkeywords={\paperkeywords},
            pdfcopyright={Copyright (C) \papercopyright},
            pdflicenseurl={\paperlicenceurl}]{hyperref}

\usepackage[colorinlistoftodos,textsize=scriptsize]{todonotes}

\usepackage[bottom,flushmargin,hang,multiple]{footmisc}

\usepackage[all]{hypcap} 

\usepackage{xspace} 
\usepackage{upgreek}

\def\lhcb   {\mbox{LHCb}\xspace}

\def\MagUp {\mbox{\em Mag\kern -0.05em Up}\xspace}

\ifthenelse{\boolean{uprightparticles}}%
{

 \def\Ppi         {\ensuremath{\uppi}\xspace}

 \def\Ppsi        {\ensuremath{\uppsi}\xspace}

 \def\PDelta      {\ensuremath{\Delta}\xspace}                 
 \def\PXi         {\ensuremath{\Xi}\xspace}                 
 \def\PLambda     {\ensuremath{\Lambda}\xspace}                 
 \def\PSigma      {\ensuremath{\Sigma}\xspace}                 
 \def\POmega      {\ensuremath{\Omega}\xspace}                 
 \def\PUpsilon    {\ensuremath{\Upsilon}\xspace}
 \let\oldPi\Pi
 \def\PPi         {\ensuremath{\oldPi}\xspace}

 \def\PB      {\ensuremath{\mathrm{B}}\xspace}                 
                  
 \def\PD      {\ensuremath{\mathrm{D}}\xspace}

 \def\PJ      {\ensuremath{\mathrm{J}}\xspace}                 
 \def\PK      {\ensuremath{\mathrm{K}}\xspace}

 \def\PW      {\ensuremath{\mathrm{W}}\xspace}

 \def\Pb      {\ensuremath{\mathrm{b}}\xspace}                 
 \def\Pc      {\ensuremath{\mathrm{c}}\xspace}

 \def\Pi      {\ensuremath{\mathrm{i}}\xspace}

 \def\Pp      {\ensuremath{\mathrm{p}}\xspace}

 \def\Ps      {\ensuremath{\mathrm{s}}\xspace}

 \def\thebaroffset{0.0em}
}
{

 \def\Ppi         {\ensuremath{\pi}\xspace}

 \def\Ppsi        {\ensuremath{\psi}\xspace}                 
                  
 \mathchardef\PDelta="7101
 \mathchardef\PXi="7104
 \mathchardef\PLambda="7103
 \mathchardef\PSigma="7106
 \mathchardef\POmega="710A
 \mathchardef\PUpsilon="7107
 \mathchardef\PPi="7105
                  
 \def\PB      {\ensuremath{B}\xspace}                 
                  
 \def\PD      {\ensuremath{D}\xspace}

 \def\PJ      {\ensuremath{J}\xspace}                 
 \def\PK      {\ensuremath{K}\xspace}

 \def\PW      {\ensuremath{W}\xspace}

 \def\Pb      {\ensuremath{b}\xspace}                 
 \def\Pc      {\ensuremath{c}\xspace}

 \def\Pi      {\ensuremath{i}\xspace}

 \def\Pp      {\ensuremath{p}\xspace}

 \def\Ps      {\ensuremath{s}\xspace}

 \def\thebaroffset{0.18em}
}
\newcommand{\offsetoverline}[2][\thebaroffset]{\kern #1\overline{\kern -#1 #2}}%

\makeatletter
\ifcase \@ptsize \relax
  \newcommand{\miniscule}{\@setfontsize\miniscule{4}{5}}
\or
  \newcommand{\miniscule}{\@setfontsize\miniscule{5}{6}}
\or
  \newcommand{\miniscule}{\@setfontsize\miniscule{5}{6}}
\fi
\makeatother

\DeclareRobustCommand{\optbar}[1]{\shortstack{{\miniscule (\rule[.5ex]{1.25em}{.18mm})}
  \\ [-.7ex] $#1$}}

\def\Wm     {{\ensuremath{\PW^-}}\xspace}

\def\squark    {{\ensuremath{\Ps}}\xspace}

\def\cquark    {{\ensuremath{\Pc}}\xspace}

\def\bquark    {{\ensuremath{\Pb}}\xspace}

\def\pion   {{\ensuremath{\Ppi}}\xspace}

\def\pip    {{\ensuremath{\pion^+}}\xspace}
\def\pim    {{\ensuremath{\pion^-}}\xspace}

\def\kaon    {{\ensuremath{\PK}}\xspace}

\def\KorKbar {\kern \thebaroffset\optbar{\kern -\thebaroffset \PK}{}\xspace}

\def\Kp      {{\ensuremath{\kaon^+}}\xspace}
\def\Km      {{\ensuremath{\kaon^-}}\xspace}

\def\Dbar    {{\ensuremath{\offsetoverline{\PD}}}\xspace}
\def\D       {{\ensuremath{\PD}}\xspace}

\def\DorDbar {\kern \thebaroffset\optbar{\kern -\thebaroffset \PD}\xspace}
\def\Dz      {{\ensuremath{\D^0}}\xspace}
\def\Dzb     {{\ensuremath{\Dbar{}^0}}\xspace}
\def\Dp      {{\ensuremath{\D^+}}\xspace}
\def\Dm      {{\ensuremath{\D^-}}\xspace}

\def\DpDm    {\ensuremath{\Dp {\kern -0.16em \Dm}}\xspace}

\def\Dstarm  {{\ensuremath{\D^{*-}}}\xspace}

\def\Ds      {{\ensuremath{\D^+_\squark}}\xspace}

\def\B       {{\ensuremath{\PB}}\xspace}

\def\BorBbar {\kern \thebaroffset\optbar{\kern -\thebaroffset \PB}\xspace}

\def\Bd      {{\ensuremath{\B^0}}\xspace}

\def\BdorBdbar {\kern \thebaroffset\optbar{\kern -\thebaroffset \Bd}\xspace}

\def\Bs      {{\ensuremath{\B^0_\squark}}\xspace}

\def\BsorBsbar {\kern \thebaroffset\optbar{\kern -\thebaroffset \Bs}\xspace}

\def\jpsi     {{\ensuremath{{\PJ\mskip -3mu/\mskip -2mu\Ppsi}}}\xspace}

\def\Y#1S{\ensuremath{\PUpsilon{(#1S)}}\xspace}

\def\proton      {{\ensuremath{\Pp}}\xspace}

\def\Lz          {{\ensuremath{\PLambda}}\xspace}

\def\LorLbar     {\kern \thebaroffset\optbar{\kern -\thebaroffset \PLambda}\xspace}

\def\Sigmares    {{\ensuremath{\PSigma}}\xspace}

\def\Lc          {{\ensuremath{\Lz^+_\cquark}}\xspace}

\def\Sigmacp      {{\ensuremath{\Sigmares_\cquark^+}}\xspace}

\def\Lb           {{\ensuremath{\Lz^0_\bquark}}\xspace}

\def\BF         {{\ensuremath{\mathcal{B}}}\xspace}
\def\BR         {\BF}

\newcommand{\decay}[2]{\ensuremath{#1\!\to #2}\xspace} 

\def\to                 {\ensuremath{\rightarrow}\xspace}

\def\AT#1     {\ensuremath{A_{\mathrm{T}}^{#1}}\xspace}           

\def\C#1      {\ensuremath{\mathcal{C}_{#1}}\xspace}                       
\def\Cp#1     {\ensuremath{\mathcal{C}_{#1}^{'}}\xspace}                    
\def\Ceff#1   {\ensuremath{\mathcal{C}_{#1}^{\mathrm{(eff)}}}\xspace}        
\def\Cpeff#1  {\ensuremath{\mathcal{C}_{#1}^{'\mathrm{(eff)}}}\xspace}       
\def\Ope#1    {\ensuremath{\mathcal{O}_{#1}}\xspace}                       
\def\Opep#1   {\ensuremath{\mathcal{O}_{#1}^{'}}\xspace}                    

\newcommand{\aunit}[1]{\ensuremath{\text{\,#1}}}       

\newcommand{\tev}{\aunit{Te\kern -0.1em V}\xspace}
\newcommand{\gev}{\aunit{Ge\kern -0.1em V}\xspace}
\newcommand{\mev}{\aunit{Me\kern -0.1em V}\xspace}
\newcommand{\kev}{\aunit{ke\kern -0.1em V}\xspace}
\newcommand{\ev}{\aunit{e\kern -0.1em V}\xspace}
 
\newcommand{\mevc}{\ensuremath{\aunit{Me\kern -0.1em V\!/}c}\xspace}
\newcommand{\gevc}{\ensuremath{\aunit{Ge\kern -0.1em V\!/}c}\xspace}
\newcommand{\mevcc}{\ensuremath{\aunit{Me\kern -0.1em V\!/}c^2}\xspace}
\newcommand{\gevcc}{\ensuremath{\aunit{Ge\kern -0.1em V\!/}c^2}\xspace}

\def\mm   {\aunit{mm}\xspace}

\def\fb   {\ensuremath{\aunit{fb}}\xspace}
\def\invfb   {\ensuremath{\fb^{-1}}\xspace}

\def\ps   {\ensuremath{\aunit{ps}}\xspace}

\def\gsim{{~\raise.15em\hbox{$>$}\kern-.85em
          \lower.35em\hbox{$\sim$}~}\xspace}
\def\lsim{{~\raise.15em\hbox{$<$}\kern-.85em
          \lower.35em\hbox{$\sim$}~}\xspace}

\def\pt         {\ensuremath{p_{\mathrm{T}}}\xspace}

\def\mrad{\aunit{mrad}\xspace}

\def\evtgen     {\mbox{\textsc{EvtGen}}\xspace}

\def\geant      {\mbox{\textsc{Geant4}}\xspace}

\def\photos     {\mbox{\textsc{Photos}}\xspace}

\def\pythia     {\mbox{\textsc{Pythia}}\xspace}

\def\tell1  {TELL1\xspace}
\def\ukl1   {UKL1\xspace}

\newcommand{\lhcborcid}[1]{\href{https://orcid.org/#1}{\hspace*{0.1em}\raisebox{-0.45ex}{\includegraphics[width=1em]{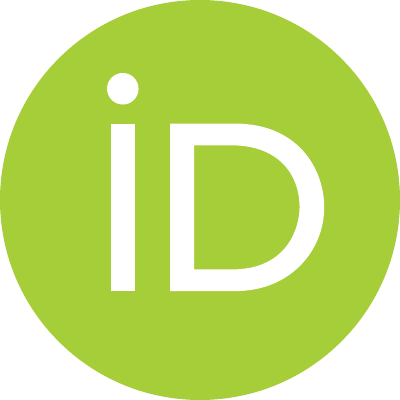}}}}

\usepackage{cite} 
\usepackage{mciteplus}

\usepackage{longtable} 
\begin{document}

\renewcommand{\thefootnote}{\fnsymbol{footnote}}
\setcounter{footnote}{1}


\begin{titlepage}
\pagenumbering{roman}

\vspace*{-1.5cm}
\centerline{\large EUROPEAN ORGANIZATION FOR NUCLEAR RESEARCH (CERN)}
\vspace*{1.5cm}
\noindent
\begin{tabular*}{\linewidth}{lc@{\extracolsep{\fill}}r@{\extracolsep{0pt}}}
\\
 & & CERN-EP-2024-098 \\  
 & & LHCb-PAPER-2023-044 \\  
 & & 22 August 2024 \\ 
 & & \\
\end{tabular*}

\vspace*{1.0cm}

{\normalfont\bfseries\boldmath\huge
\begin{center}
\begin{spacing}{1.25}
\papertitle 
\end{spacing}
\end{center}
}

\vspace*{1.7cm}

\begin{center}
\paperauthors\footnote{Full author list given at the end of the article.}
\end{center}

\vspace{\fill}



\begin{abstract}
  \noindent
  The four decays, $\Lbdecays$, are observed for the first time using proton-proton collision data collected with the \lhcb detector at a center-of-mass energy of $13\tev$, corresponding to an integrated luminosity of $6\invfb$. By considering the \LbtoLcDzK decay as reference channel, the following branching fraction ratios are measured to be
	\begin{equation*}
		\begin{split}
            \frac{\BF\left(\LbtoScDmK\right)}{\BF\left(\LbtoLcDzK\right)}
            & = \RBRresult{0.282}{0.016}{0.016}{0.005},\\
            \frac{\BR(\LbtoScstDmK)}{\BR(\LbtoScDmK)}
            & = \RBRresultNoBrErr{0.460}{0.052}{0.028},\\
            \frac{\BR(\LbtoScDstK)}{\BR(\LbtoScDmK)}
            & = \RBRresult{2.261}{0.202}{0.129}{0.046},\\
            \frac{\BR(\LbtoScstDstK)}{\BR(\LbtoScDmK)}
            & = \RBRresult{0.896}{0.137}{0.066}{0.018},
		\end{split}
	\end{equation*}
	where the first uncertainties are statistical, the second are systematic, and the third are due to uncertainties in the branching fractions of intermediate particle decays. These initial observations mark the beginning of pentaquark searches in these modes, with more datasets to become available following the LHCb upgrade.
\end{abstract}

\vspace*{1.4cm}

\begin{center}
  Published in Phys. Rev. D 110, L031104
\end{center}

\vspace{\fill}

{\footnotesize 
\centerline{\copyright~\papercopyright. \href{\paperlicenceurl}{\paperlicence}.}}
\vspace*{2mm}

\end{titlepage}


\newpage
\setcounter{page}{2}
\mbox{~}
%
%
%
%


\renewcommand{\thefootnote}{\arabic{footnote}}
\setcounter{footnote}{0}

\cleardoublepage


\pagestyle{plain} 
\setcounter{page}{1}
\pagenumbering{arabic}


\section{Introduction}
The existence of exotic pentaquark states, comprising four quarks and an antiquark, has been predicted since the establishment of the quark model~\cite{GellMann:1964nj,*Zweig:352337}.
The search for pentaquark candidates has been performed by many experiments in the past 50 years~\cite{Pennington:2016dpj,Amaryan:2022iij,Rossi:2020ezg} but only the LHCb experiment has given conclusive results.
In 2015, the LHCb experiment reported the observation of $\jpsi p$ resonant structures~\cite{LHCb-PAPER-2015-029,LHCb-PAPER-2016-009}, consistent with pentaquark candidates made up of minimal quark content $c\bar{c}uud$, produced in $\Lb\to\jpsi p \Km$ decays.\footnote{Inclusion of charge-conjugate processes is implied throughout.}
An amplitude analysis showed that the data is best described with the inclusion of two pentaquark states, the $P_{\psi}^{N}(4380)^+$ and $P_{\psi}^{N}(4450)^+$.
With additional data and an improved selection strategy, it was found that the $P_{\psi}^{N}(4450)^+$ structure resolves into two narrower substructures, the $P_{\psi}^{N}(4440)^+$ and $P_{\psi}^{N}(4457)^+$.
In addition, a new narrow pentaquark candidate, the $P_{\psi}^{N}(4312)^+$ was discovered~\cite{LHCb-PAPER-2019-014}. 

The newly observed exotic candidates have masses less than $10\mev$ below the $\PSigma_c \DorDstarb$ thresholds.\footnote{Natural units with $\hbar = c= 1$ are used throughout.}
The proximity of the pentaquark candidates to open-charm thresholds, as well as their very narrow widths, favor the loosely bound (so-called molecular) pentaquark model, in which the $\DorDstarb$ meson and the $\PSigma_{c}^{(*)}$ baryon are bound by a residual strong force similar to that binding a proton and neutron to form a deuteron~\cite{Weinberg:Deuteron1965}.
In addition, several molecular pentaquark models predict that a ${P_{\psi}^{N+}}$ with $3/2^-$ spin parity would decay substantially into $\PSigma_{c}^{(*)}\Dbar$~\cite{Du:2021fmf,Burns:2021jlu,He:2019rva,Pan:2023hrk}.
This motivates the search for $\Lbdecays$ decays,\footnote{The symbols \Sigmacpp and \Sigmacstarpp refer to $\PSigma_c(2455)^{++}$ and $\PSigma_c(2520)^{++}$.} whose tree-level Feynman diagrams, shown in Fig.~\ref{fig:feynmandiag} (left), are color suppressed but not forbidden~\cite{Liu:2015fea}.
The topologically similar $\LbtoLcDzK$ process is chosen as the reference mode, which has contributions from both color-suppressed and color-favored tree-level diagrams.
As shown in Fig.~\ref{fig:feynmandiag} (right) for the color-favored process, the spectator quarks ($ud$) of the $\Lb$ baryon directly propagate into the charmed baryon ($udc$), which must preserve the isospin-0 quantum number of the parent $\Lb$ baryon, and thus be a $\Lc$ state. 

\begin{figure}[b]
\centering
\scalebox{0.85}{
\begin{tikzpicture}
\begin{feynman}
\vertex (a1) {\(u [d]\)};
\vertex[right=2.0cm of a1] (a2);
\vertex[right=3.0cm of a2] (a3) {\(u [d]\)};
\vertex[below=1.5em of a1] (b1) {\(\,\,\,\,\,\,\,b   \)};
\vertex[below=1.5em of a3] (b3) {\(c\,\,\,\,\,\,\,    \)};
\vertex[below=1.5em of a2] (b2);
\vertex[below=1.5em of b3] (c1) {\(u\,\,\,\,\,\,\,    \)};
\vertex[below=2.5em of c1] (c3) {\(\bar{u}\,\,\,\,\,\, \,   \)};
\vertex[below=9.5em of b1] (d1) {\(d [u]\)};
\vertex[below=9.5em of b3] (d3) {\(d [u]\)};
\vertex[below=9.5em of b2] (d2);
\vertex[below=1.5em of c3] (e1) {\(s\,\,\,\,\,\,\,    \)};
\vertex[below=2.5em of e1] (e3) {\(\bar{c}\,\,\,\,\,\,\,    \)};
\vertex at ($(c1)!0.5!(c3) - (1.5cm, 0)$) (c2);
\vertex at ($(e1)!0.5!(e3) - (1.8cm, 0)$) (e2);
\diagram* {
(a1) -- [fermion] (a2),
(a2) -- [fermion] (a3),
(b1) -- [fermion] (b2),
(b2) -- [fermion] (b3),
(d1) -- [fermion] (d2),
(d2) -- [fermion] (d3),
(c3) -- [fermion, out=180, in=-30] (c2) -- [fermion, out=30, in=180] (c1),
(e3) -- [fermion, out=180, in=-30] (e2) -- [fermion, out=30, in=180] (e1),
(b2) -- [boson,  edge label'=\(\Wm\)] (e2),
};
\draw [decoration={brace}, decorate] (d1.south west) -- (a1.north west)
node [pos=0.5, left=0.1cm] {\(\Lb \)};
\draw [decoration={brace}, decorate] (a3.north east) -- (c1.south east)
node [pos=0.5, right=0.1cm, align=left] {\(\scstpp \,\,\,\,\,\, \,\,\,\,\,\, \) \\ \([\Lc/\Sigmacp]\)};
\draw [decoration={brace}, decorate] (c3.north east) -- (e1.south east)
node [pos=0.5, right=0.1cm] {\(\Km\)};
\draw [decoration={brace}, decorate] (e3.north east) -- (d3.south east)
node [pos=0.5, right=0.1cm, align=left] {\(\DorDstarm \) \\ \([\DorDstarbz]\)};

\end{feynman}
\end{tikzpicture}

}
\scalebox{0.85}{
\begin{tikzpicture}
\begin{feynman}
\vertex (a1) {\(d\)};
\vertex[right=2.0cm of a1] (a2);
\vertex[right=3.0cm of a2] (a3) {\(d\)};
\vertex[below=1.5em of a1] (d1) {\(u\)};
\vertex[below=1.5em of a3] (d3) {\(u\)};
\vertex[below=1.5em of a2] (d2);
\vertex[below=1.5em of d1] (b1) {\(b\)};
\vertex[below=1.5em of d3] (b3) {\(c\)};
\vertex[below=1.5em of d2] (b2);
\vertex[below=2.0em of b3] (c1) {\(s\)};
\vertex[below=1.5em of c1] (e3) {\(\bar{u}\)};
\vertex[below=3.0em of e3] (e1) {\(u\)};
\vertex[below=1.5em of e1] (c3) {\(\bar{c}\)};
\vertex at ($(c1)!0.5!(c3) - (1.8cm, 0)$) (c2);
\vertex at ($(e1)!0.5!(e3) - (1.2cm, 0)$) (e2);
\diagram* {
(a1) -- [fermion] (a2),
(a2) -- [fermion] (a3),
(b1) -- [fermion] (b2),
(b2) -- [fermion] (b3),
(d1) -- [fermion] (d2),
(d2) -- [fermion] (d3),
(c3) -- [fermion, out=180, in=-50] (c2) -- [fermion, out=50, in=180] (c1),
(e3) -- [fermion, out=180, in=30] (e2) -- [fermion, out=-30, in=180] (e1),
(b2) -- [boson,  edge label'=\(\Wm\)] (c2),
};
\draw [decoration={brace}, decorate] (b1.south west) -- (a1.north west)
node [pos=0.5, left=0.1cm] {\(\Lb \)};
\draw [decoration={brace}, decorate] (a3.north east) -- (b3.south east)
node [pos=0.5, right=0.1cm] {\(\Lc\)};
\draw [decoration={brace}, decorate] (e1.north east) -- (c3.south east)
node [pos=0.5, right=0.1cm] {\(\DorDstarbz\)};
\draw [decoration={brace}, decorate] (c1.north east) -- (e3.south east)
node [pos=0.5, right=0.1cm] {\(\Km\)};

\end{feynman}
\end{tikzpicture}
}
\caption{Feynman diagrams for (left) color-suppressed and (right) color-favored tree processes of the \Lb baryon decaying into $\PSigma_{c}^{(*)} \DorDstarb \Km$ or $\Lc \DorDstarbz \Km$.}
\label{fig:feynmandiag}
\end{figure}

This article presents the search for four new $\Lbdecays$ decay modes using proton-proton ($pp$) collision data collected by the \lhcb experiment at a center-of-mass energy of $13\tev$ during the Run 2 data-taking period from 2015 to 2018, corresponding to an integrated
luminosity of $6\invfb$.
The relative branching fractions, 	$\frac{\BF\left(\LbtoScDmK\right)}{\BF\left(\LbtoLcDzK\right)}$, $\frac{\BR(\LbtoScstDmK)}{\BR(\LbtoScDmK)}$, $	\frac{\BR(\LbtoScDstK)}{\BR(\LbtoScDmK)}$, and $\frac{\BR(\LbtoScstDstK)}{\BR(\LbtoScDmK)}$ are measured. 
Given that the decay modes in the last three ratios have the same number of final tracks and final states similar to those of the \LbtoScDmK decay mode, their relative branching fractions with respect to the \LbtoScDmK decay are measured to facilitate the cancellation of several systematic uncertainties due to the effects from detector acceptance, triggers, and tracking.

\section{Detector and simulation}


The LHCb 
detector~\cite{LHCb-DP-2008-001,LHCb-DP-2014-002} is a
single-arm forward spectrometer covering the pseudorapidity range $2 < \eta < 5$, designed for
the study of particles containing \bquark\ or \cquark\ quarks. The detector elements that are particularly
relevant to this analysis are a silicon-strip vertex detector surrounding the $pp$ interaction
region that allows \cquark\ and \bquark\ hadrons to be identified from their characteristically long
flight distance; a tracking system that provides a measurement of the momentum, $p$, of charged
particles; and two ring-imaging Cherenkov detectors that are able to discriminate between
different species of charged hadrons.

The online event selection is performed by a trigger~\cite{LHCb-DP-2012-004}, which consists of a hardware and a software stage. 
At the hardware stage, events are required to have a muon with high transverse momentum, \pt, or a hadron, photon, or electron with high transverse energy in the calorimeters.
The triggered objects can be either hadrons from the \Lb decays of interest [trigger on signal, (TOS)] or any particle from the rest of the event [trigger independently of signal, (TIS)].
The software trigger applies a full event reconstruction, and subsequently requires at least one charged particle from the event to have a larger \pt and be inconsistent with originating from any reconstructed primary $pp$ interaction vertex (PV).
Furthermore, the reconstructed events must have signal candidates forming a two-, three-, or four-track secondary vertex significantly displaced from any PV.
The secondary vertices are filtered by a multivariate algorithm~\cite{LHCb-PROC-2015-018} to be consistent with the decay of a beauty hadron.

Simulation is required to model the resolution effects of event reconstruction and to calculate the efficiencies due to detector acceptance and selection requirements.
In the simulation, $pp$ collisions are generated 
using \pythia~\cite{Sjostrand:2007gs,*Sjostrand:2006za} with a specific \lhcb configuration~\cite{LHCb-PROC-2010-056}.
Decays of unstable particles are described by \evtgen~\cite{Lange:2001uf}, in which final-state radiation is generated using \photos~\cite{davidson2015photos}.
The interaction of the generated particles with the detector, and its response, are implemented using the \geant toolkit~\cite{Allison:2006ve, *Agostinelli:2002hh} as described in Ref.~\cite{LHCb-PROC-2011-006}.
The particle-identification (PID) response is not well described in the LHCb simulation and is corrected to match that in data using dedicated calibration samples.
The corrections are based on a four-dimensional kernel density estimation for distributions in the PID response, $\pt$ and $\eta$ of the track, and the multiplicity of the event~\cite{Poluektov:2014rxa}.

\section{Candidate selection}

The \Lbdecays candidates are reconstructed through the decay chains \mbox{$\scstpp\to \Lc \pip$} with \mbox{$\Lc\to p \Km \pip$} and \mbox{$\Dm \to \Kp \pim \pim$} or \mbox{$\Dstarm  \to \Dzb \pim$} with \mbox{$\Dzb\to \Kp \pim$}.
The reconstructed final-state particles are required to have PID information consistent with their respective mass hypotheses and be inconsistent with originating from any PV. 
At least one of them must have $p > 10 \gev$ and $\pt > 1.7\gev$. 
Each of the $\Lc$, $\Dzb$, and $\DorDstarm$ decay vertices must have a good vertex-fit quality and be significantly displaced from its associated PV, defined as the primary vertex that fits best to the flight direction of the candidate. 
The reconstructed masses of the charmed hadron candidates must be consistent with their known values~\cite{PDG2022}, within $15\mev$ for the $\Lc$ baryon and $25\mev$ for the charmed mesons. 
Subsequently, the $\Lb$ candidate is reconstructed by combining $\Lc$, $\DorDstarm$, $K^-$ and $\pip$ candidates to form a good vertex.
The momentum vector of the \Lb candidate is required to be consistent with the flight direction.
The sum of transverse momentum of the $\Lb$ decay products must be greater than $5\gev$ and the decay time of the $\Lb$ candidate is required to be greater than $0.2\ps$.
Clone tracks are pairs of reconstructed tracks that share a majority of their detector hits.
They are rejected by requiring the opening angle between any pair of final-state tracks from the $\Lb$ candidate to be greater than $0.5\mrad$.
The reference channel $\LbtoLcDzK$ is reconstructed from the same $\Lc$ and $\Dzb$ decay channels by applying the same selections.

Non-negligible peaking backgrounds in the signal and reference channels result from $\Lb$ decays to the same final state but without one or both intermediate charm hadrons. These are denoted by non-doubly-charmed (NDC) backgrounds. 
Genuine \Lc and \Dzb(\Dm) hadrons have nonzero lifetimes and decay at tertiary vertices in the forward direction with respect to the \Lb baryon decay vertex.
To suppress this background, the separation along the beam direction $z$ between the charmed hadrons and the \Lb decay vertex is required to be greater than $-1\mm$ and its significance must be greater than $1$, $-2.5$, $-1.5$, and $-2.0$ for \Dm in signal decay, \Lc in signal decay, \Dzb in reference decay, and \Lc in reference decay, respectively. 
The negative values of these requirements account for the limited resolution of the vertex reconstruction.

Specific backgrounds from misreconstructed or misidentified particles are vetoed by applying selections on the invariant masses.
A misreconstructed background can arise from the two final-state kaons being correctly identified but assigned to the wrong \Lc and \Lb parent particles.
Therefore, peaks in \mbox{$\Lc\to(p\pip)_\Lc K^-_{\Lb}$} are vetoed, where the subscript in each final state particle denotes the assumed parent during reconstruction.

For misidentified backgrounds, the vetoes are applied by assigning an alternative mass hypothesis to the final state particles, and rejecting candidates that have a recalculated invariant mass consistent with known resonances.
Misidentified background vetoes are applied for \mbox{$\Ds\to(\{\Kp\Rightarrow p\}\Km\pip)_\Lc$}, \mbox{$\phi\to(\{\Kp\Rightarrow p\}\Km)_\Lc$}, \mbox{$\phi\to \{\Kp\Rightarrow p\}_\Lc K^-_{\Lb}$}, \mbox{$\Dz\to (\Km \{\pip\Rightarrow p\})_\Lc$}, \mbox{$\Lc\to (p \pip)_\Lc \{\Km\Rightarrow\pim\}_{\Dzb/\Dm}$} and \mbox{$\Dstarm\to\Dzb_\Lb\{\pi^-\Rightarrow\Km\}_{\Lb/\Lc}$} decays\footnote{The particle species on the left and right of the arrow ($\Rightarrow$) correspond to the alternative and default mass hypotheses, respectively.}, where the subscript denotes the parent assumed during reconstruction. The last veto is only applied to decay chains with a $\Dzb$ meson, while the others are applied to all decay modes.

After applying all selections, a few percent of events contain multiple candidates.
These are mostly due to duplicated use of the same tracks. 
In such events, the candidate with the smallest value of $\chi^2$ from a kinematic fit with constrained decay vertices is retained. 

\section{Mass fit}

The yields of the \mbox{\LbtoSDmK} (\mbox{\LbtoSDstK}) signal decays are determined from unbinned two-dimensional maximum likelihood fits to the \mbox{$m_{\Lc\pip}$} and \mbox{$m_{\Lc\Dm\Km\pip}$} (\mbox{$m_{\Lc\Dstarm\Km\pip}$}) invariant mass distributions, where the subscript denotes the particle combination used to calculate the invariant masses.
The \mbox{$m_{\Lc\pip}$} dimension is used to disentangle the contributions from \Sigmacpp, \Sigmacstarpp and nonresonant \mbox{$\Lc\pip$}.
To reduce the correlation between \mbox{$m_{\Lc\pip}$} and \mbox{$m_{\Lc\Dm\Km\pip}$} (\mbox{$m_{\Lc\Dstarm\Km\pip}$}) and improve the resolution of their mass spectra, the invariant mass \mbox{$m_{\Lc\pip}$} is reconstructed by a kinematic fit which requires the \Lb to originate from its associated PV and constrains the \Lb, \Lc and \Dm(\Dstarm) to their known masses\cite{PDG2022}.
The same PV constraint is also applied when reconstructing \mbox{$m_{\Lc\Dm\Km\pip}$} (\mbox{$m_{\Lc\Dstarm\Km\pip}$}) but only the \Lc and \Dm(\Dstarm) masses are constrained.

Both two-dimensional fits contain six components each.
The signal decays \mbox{\LbtoSDmK} and \mbox{\LbtoSDstK} include a resonant \scstpp and a peaking \Lb component.
The resonant \scstpp distribution is modeled by an incoherent relativistic Breit--Wigner distribution convolved with a Gaussian resolution function.
The Breit--Wigner masses and widths of the \scstpp state are fixed to their known values\cite{PDG2022}, while the detector resolution is obtained from simulation.
The \Lb mass peak is modeled by the sum of two Crystal Ball~\cite{Skwarnicki:1986xj} functions, one with a low-mass tail and the other with a high-mass tail.
The two functions share a common mean.
For each decay mode, the shape parameters are determined from fits to simulation, except for the mean and a width scale factor which are free parameters in the data fit to account for imperfections in the simulation. 

A purely combinatorial background component is described by a threshold function in $m_{\Lc\pip}$ and an exponential function in $m_{\Lc\Dm\Km\pip}$ ($m_{\Lc\Dstarm\Km\pip}$), whose parameters are determined from data.
Two background components with resonant \scstpp states but nonpeaking in the \Lb invariant mass share the \scstpp shape parameters with the signal modes and share the $m_{\Lc\Dm\Km\pip}$ ($m_{\Lc\Dstarm\Km\pip}$) exponential slope with each other, but not with the pure combinatorial background.
A peaking \Lb background component with nonresonant \Lc\pip shares the peaking \Lb shape parameters with the signal modes and shares the $m_{\Lc\pip}$ threshold function parameters with the pure combinatorial background.

\begin{figure}[!bt]
    \centering
    \captionsetup[subfigure]{labelformat=empty}
    \begin{subfigure}{0.48\textwidth}
        \centering
        \includegraphics[width=\textwidth]{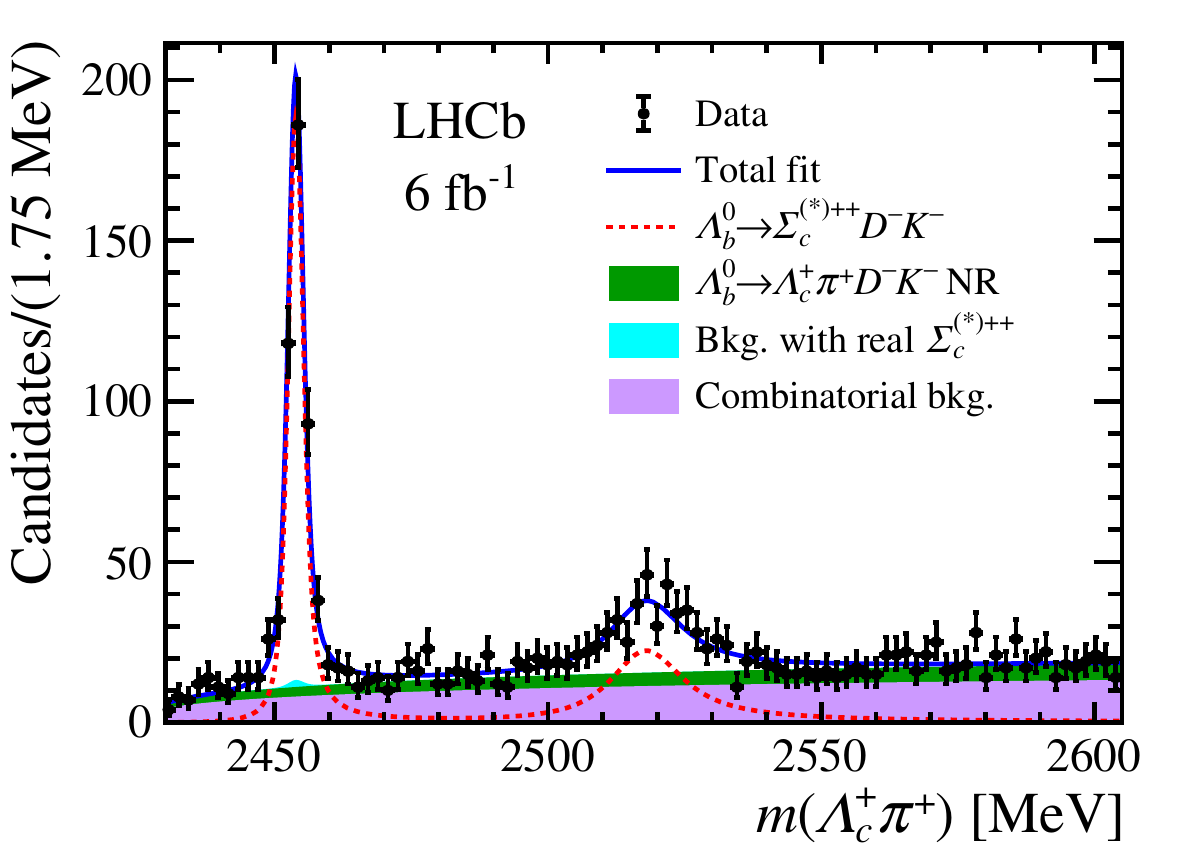}
        \label{fig:2DFit_ScDK_Sc}
    \end{subfigure}
    \begin{subfigure}{0.48\textwidth}
        \centering
        \includegraphics[width=\textwidth]{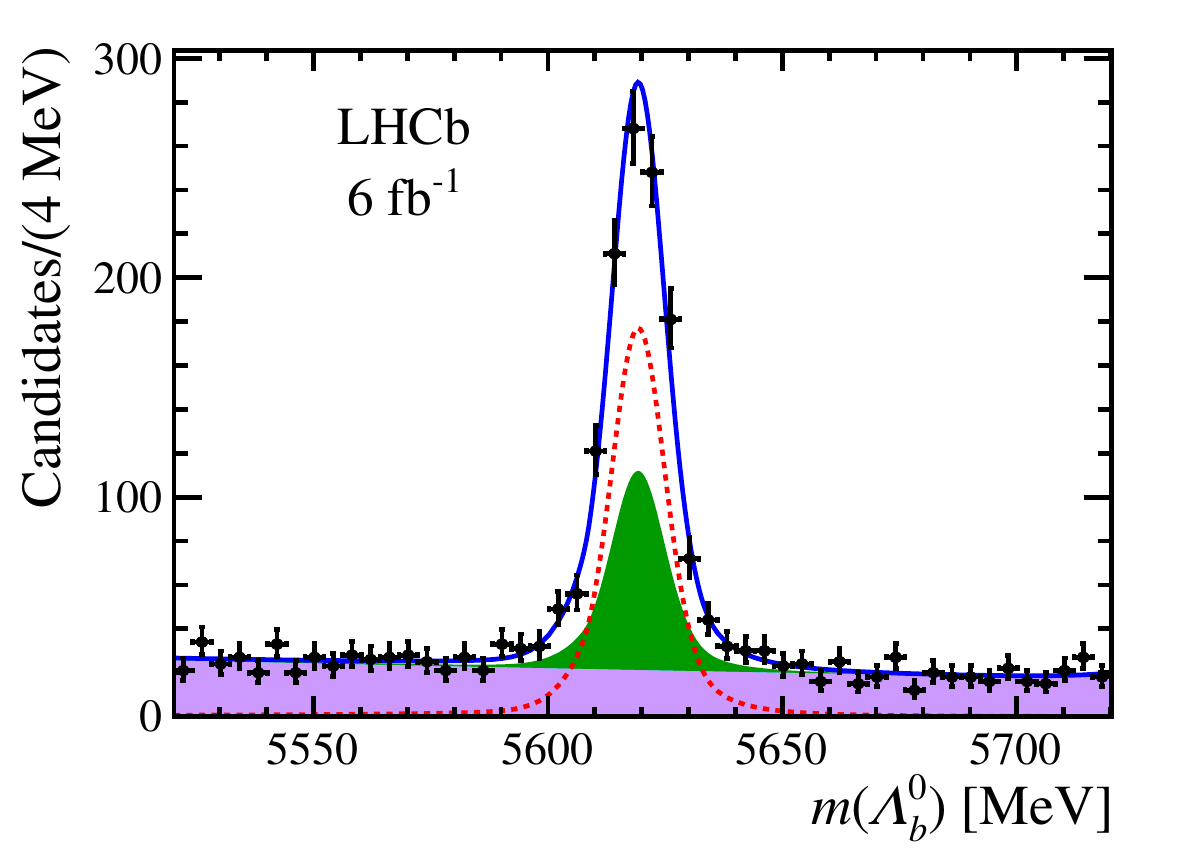}
        \label{fig:2DFit_ScDK_Lb}
    \end{subfigure}
    \begin{subfigure}{0.48\textwidth}
        \centering
        \includegraphics[width=\textwidth]{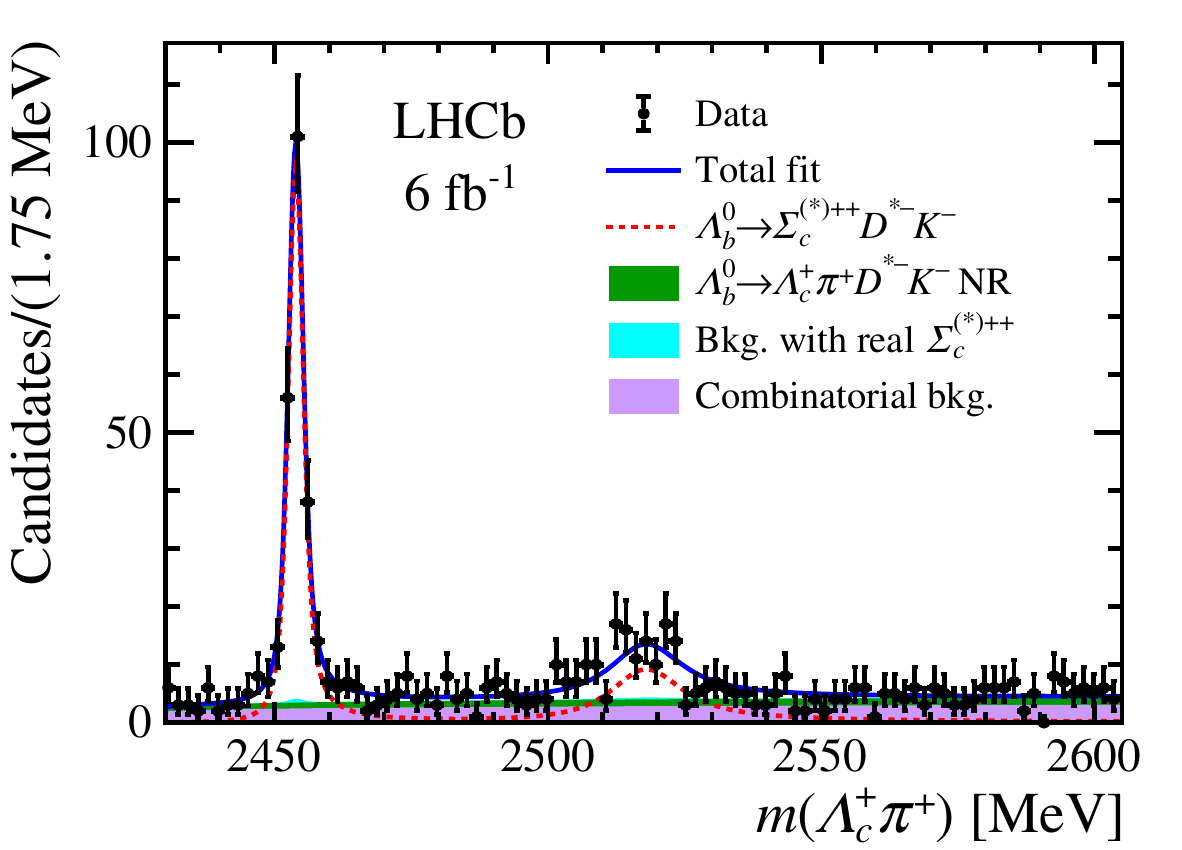}
        \label{fig:2DFit_ScDstK_Sc}
    \end{subfigure}
    \begin{subfigure}{0.48\textwidth}
        \centering
        \includegraphics[width=\textwidth]{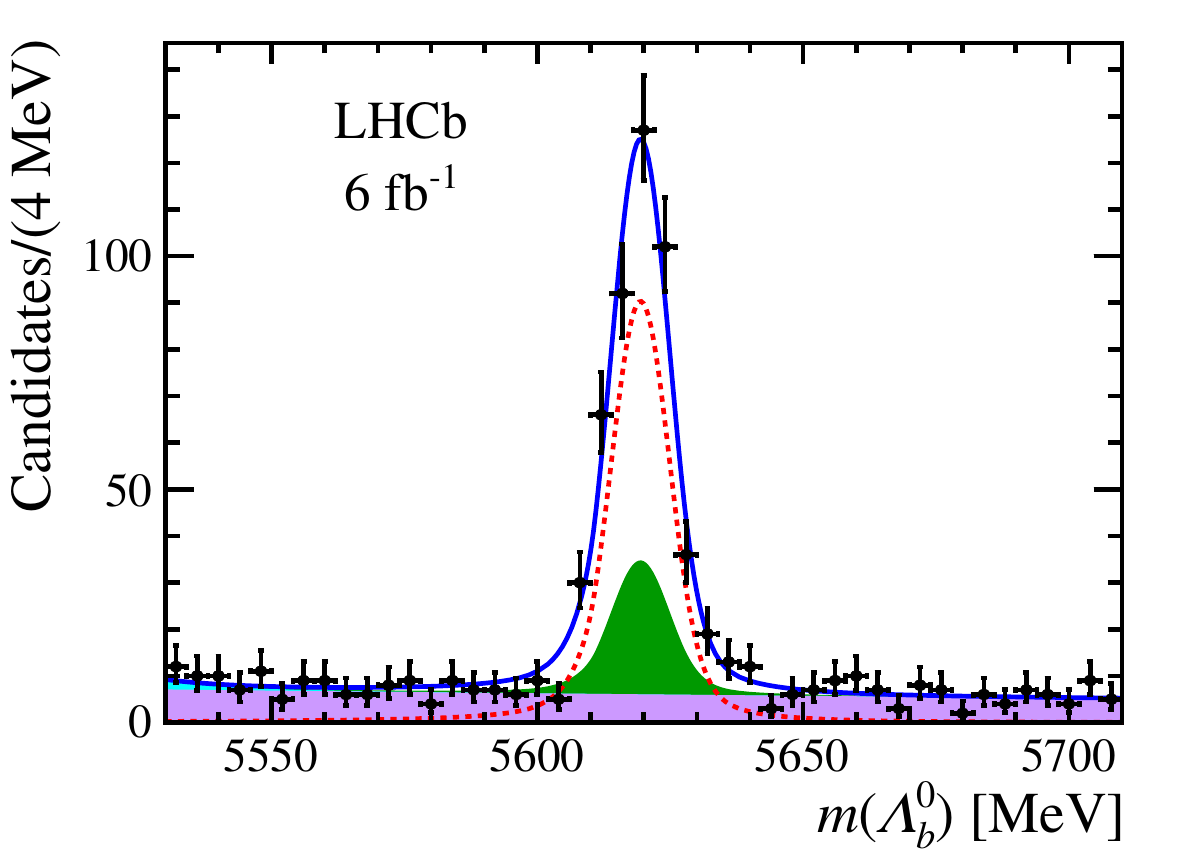}
        \label{fig:2DFit_ScDstK_Lb}
    \end{subfigure}
    \caption{Two-dimensional invariant mass fits of the \LbtoSDmK decay, projected onto (top left) \mbox{$m_{\Lc\pip}$} and (top right) \mbox{$m_{\Lc\Dm\Km\pip}$}. A similar fit to \LbtoSDstK decay is projected onto (bottom left) \mbox{$m_{\Lc\pip}$} and (bottom right) \mbox{$m_{\Lc\Dstarm\Km\pip}$}. The two signal contributions with resonant \Sigmacpp and \Sigmacstarpp are drawn as a single component (red dashed line). Similarly, the two resonant \scstpp backgrounds without \Lb peaks are drawn together (cyan fill).}
    \label{fig:2Dfit}
\end{figure}

The one-dimensional projections of both fits are shown in Fig.~\ref{fig:2Dfit}.
The obtained yields are $480\pm25$ for \LbtoScDmK, $279\pm26$ for \LbtoScstDmK, $243\pm17$ for \LbtoScDstK and $116\pm15$ for \LbtoScstDstK signal decays.
These results are tested for stability by generating and fitting 3000 pseudoexperiments with the default model described above, and no significant bias was observed.

The yield of the \mbox{\LbtoLcDzK} reference mode is determined by a one-dimensional fit to $m_{\Lc\Dzb\Km}$.
Similar to the treatment of the signal decay modes, \mbox{$m_{\Lc\Dzb\Km}$} is reconstructed by constraining the \Lb candidate to originate from its associated PV and applying mass constraints to the intermediate \Lc and \Dzb particles to improve the mass resolution.
For both signal and background components, the \Lb candidate spectrum is modeled in the same way as for the signal. The fitted yield of the \mbox{\LbtoLcDzK} reference decay is $4032\pm75$.

\section{Ratio of branching fractions}

The yields of the signal and reference modes are corrected for the Dalitz distribution of the \Lb decay by considering per-event efficiencies,
\begin{equation}
    N_{\text{corr}} = \sum_{i}\frac{s\mathcal{W}_{i}}{\epsilon(s^{12}_{i},s^{13}_{i})}
\end{equation}
where $N_{\text{corr}}$ is the corrected yield, $s\mathcal{W}_{i}$ is the per-event signal weight from the \splot method~\cite{Pivk:2004ty} and the efficiency, $\epsilon(s^{12}_{i},s^{13}_{i})$, is calculated event-by-event based on the Dalitz variables $s^{12}$ and $s^{13}$. These Dalitz variables represent the square of the invariant mass of $\scstpp D^{(*)-}$ system and $\scstpp\Km$ system in the three-body decay of the \Lb baryon. The efficiency is obtained using \Lb decays with uniform distribution over the phase space and applying the candidate selection.

Given the corrected yields $N_{\text{corr}}$, the ratios of branching fractions are calculated via
\begin{align}
    \frac{\mathcal{B}(\LbtoScDmK)}{\mathcal{B}(\LbtoLcDzK)} = & \frac{N_{\text{corr}}(\LbtoScDmK)}{N_{\text{corr}}(\LbtoLcDzK)}  \cdot \frac{\mathcal{B}(\DzbtoKpi)}{\mathcal{B}(\DmtoKpipi)}, \\
    \frac{\mathcal{B}(\LbtoScstDmK)}{\mathcal{B}(\LbtoScDmK)} = & \frac{N_{\text{corr}}(\LbtoScstDmK)}{N_{\text{corr}}(\LbtoScDmK)},\\
    \frac{\mathcal{B}(\LbtoScDstK)}{\mathcal{B}(\LbtoScDmK)} = & \frac{N_{\text{corr}}(\LbtoScDstK)}{N_{\text{corr}}(\LbtoScDmK)} \cdot \frac{\mathcal{B}(\DmtoKpipi)}{\mathcal{B}(\DstmtoDzbpi)\mathcal{B}(\DzbtoKpi)},\\
    \frac{\mathcal{B}(\LbtoScstDstK)}{\mathcal{B}(\LbtoScDmK)} = & \frac{N_{\text{corr}}(\LbtoScstDstK)}{N_{\text{corr}}(\LbtoScDmK)} \cdot \frac{\mathcal{B}(\DmtoKpipi)}{\mathcal{B}(\DstmtoDzbpi)\mathcal{B}(\DzbtoKpi)},
\end{align}
where the branching fractions ($\mathcal{B}$) of charmed mesons assume the values published in Ref.~\cite{PDG2022}.
The $\scstpp\to \Lc \pip$ decays are not considered in the formulas above because they are the only strong processes allowed by the mass threshold limit, and their branching fractions are assumed to be unity.

\section{Systematic uncertainties}

The four signal decay channels have the same final state and similar decay topologies, hence the systematic uncertainties from the modeling of track reconstruction efficiency in the simulation are assumed to cancel out in the efficiency ratios.
This cancellation does not apply between \mbox{\LbtoLcDzK} (reference) and \mbox{\LbtoScDmK} (signal) decays because the latter has two extra pions and an intermediate \Sigmacpp.
The modeling of the track reconstruction efficiency in simulation results in a systematic uncertainty of $1.61\%$ for each extra pion in the \LbtoScDmK decays due to the imperfect simulation of hadronic interactions with the detector material~\cite{LHCb-DP-2013-002}.

Similarly, imperfections in hardware trigger efficiencies between signal modes are assumed to cancel out in the efficiency ratios, but not in the ratio between \mbox{\LbtoLcDzK} and \mbox{\LbtoScDmK} decays.
To correct for this, the TISTOS method~\cite{LHCb-DP-2012-004} is used to derive data-driven corrections to the simulated values.
Such factors are binned in maximum \pt of the final state particles, for both the \mbox{\LbtoLcDzK} and \mbox{\LbtoScDmK} decays.
By default, the correction table derived from \LbtoLcDzK decays is used to correct the \LbtoLcDzK simulation, and likewise when correcting the \LbtoScDmK simulation.
As a systematic, only a single correction table is used to correct both decay modes, or the correction factors derived from \LbtoLcDzK and \LbtoScDmK decays are used to correct \LbtoScDmK and \LbtoLcDzK simulation respectively.
The largest change in the efficiency ratio between \LbtoLcDzK and \LbtoScDmK decays is $0.77\%$, which is assigned as the systematic uncertainty due to the hardware trigger efficiency correction.

The simulation PID response correction has two sources of uncertainty, one from the kernel density estimation and one from the finite size of the calibration samples. The width of the kernel is increased by $50\%$ and the variation in the efficiency ratios is assigned as the systematic uncertainty. For the finite size of the PID calibration sample, a bootstrapping method\cite{LHCb-PUB-2016-021} finds a change of less than $0.02\%$ in the efficiency ratios which is thus neglected.

Systematic uncertainties due to the mass fit model are estimated by using alternative signal and background probability density functions (PDFs). 
There are two alternative \scstpp Breit--Wigner PDFs: either the Blatt--Weiskopf form factor barrier radius~\cite{Blatt:1952ije}, $d$, is doubled, or the \scstpp mass and width are floated while the $m_{\Sigmacstarpp} - m_{\Sigmacpp}$ mass difference is fixed. 
The two alternative peaking \Lb PDFs are: either the fixed tail parameters are varied by $\pm 1\sigma$ of their simulation fit uncertainties or a Hypatia function~\cite{Santos:2013gra} is used instead of the two Crystal Ball functions. 
The alternative background parametrization in $m_{\Lb}$ uses a second-order Chebychev polynomial instead of the default exponential function. 
For the threshold function background PDFs in $m_{\Lc\pip}$, an additional quadratic term is multiplied to the default threshold function as an alternative PDF.
The six alternative PDFs are used one-at-a-time and two-at-a-time as alternative models to fit data, where the latter accounts for the correlation between alternative PDFs. 
The largest difference in fitted yields between any two models, default or alternative, is taken as the systematic uncertainty on the fitted yields. 
These uncertainties are subsequently propagated to the ratio of branching fractions by accounting for the correlation between different decay modes.

When calculating the default efficiencies, the kinematic distributions of the simulated \Lb particles are corrected in bins of $\eta$ and \pt to better match those in data using \mbox{\LbtoLcDzK} decays.
The systematic uncertainty of this correction is estimated by using an alternative (narrower) binning scheme.
A systematic uncertainty of $0.05\%$ is estimated for the ratio of branching fractions \mbox{$\mathcal{B}(\LbtoScDmK)/\mathcal{B}(\LbtoLcDzK)$} and is negligible for the ratios of branching fractions between signal modes.

The systematic uncertainty due to the limited size of the simulated signal and reference modes is estimated by assuming that the efficiencies follow a binomial distribution. 
The systematic uncertainty due to multiple candidate removal is calculated by assuming all removed candidates would increase the signal or reference mode yields by the number of candidates removed.
The ratios of branching fractions are recalculated with the increased yields, and the differences are assigned as a systematic uncertainty.

Although the simulated decay $\Lc \to \proton\Km\pip$ includes intermediate resonances, its simulated Dalitz distribution does not perfectly match that of real data.
The \Lc Dalitz distribution correction is estimated from \LbtoLcDzK decays by comparing the \Lc Dalitz distribution in data to simulation.
This correction is then applied to the simulated signal decay modes, and the differences in the ratios of branching fractions are taken as systematic uncertainties. 

The contamination by NDC decays of \Lb candidates can be estimated by fitting the $m_{\Lb}$ distribution in the sideband regions of $m_{\proton\Km\pip}$ and $m_{\Kp\pim}$ ($m_{\Kp\pim\pim}$), then extrapolating the yields of NDC \Lb to the \Lc and \Dm (or \Dstarm) signal window. 
This procedure estimates an NDC contamination of $0.9\%$ for \LbtoScDmK, $4.8\%$ for \LbtoScstDmK, $1.1\%$ for \LbtoSDstK, $4.1\%$ for \LbtoScstDstK and $3.2\%$ for \LbtoLcDzK.
However, the \Lc and \Dm (or \Dstarm) mass constraints smear the $m_{\Lb}$ mass peaks of NDC backgrounds reducing the potential bias that NDC backgrounds would have on the yields of doubly charmed decay modes.
As a conservative estimate, half of the NDC contamination rate is taken as a systematic uncertainty on the fitted yields, which is then propagated to the ratios of branching fractions. 

A summary of the systematic uncertainties is shown in Table~\ref{tab:syst_total}, and the systematic uncertainty for each ratio is calculated by considering the correlations between statistical limitations of simulated samples and kinematic reweighting or \Lc decay amplitude modeling, utilizing a correlation matrix with nondiagonal elements set to 1 for these sources.

\begin{table}[tb]
\centering
\caption{Summary of systematic uncertainties. The correlation between those due to the limited statistics of simulated samples and other sources is considered. Systematics uncertainties due to the modeling of track reconstruction and trigger efficiencies are assumed to cancel out in the ratios between signal modes.}
\label{tab:syst_total}
\resizebox{\columnwidth}{!}{%
\begin{tabular}{lcccc} \hline 
Source                      & $\frac{\BF\left(\LbtoScDmK\right)}{\BF\left(\LbtoLcDzK\right)}$ & $\frac{\BR(\LbtoScstDmK)}{\BR(\LbtoScDmK)}$ & $\frac{\BR(\LbtoScDstK)}{\BR(\LbtoScDmK)}$ & $\frac{\BR(\LbtoScstDstK)}{\BR(\LbtoScDmK)}$  \\ \hline
Track reconstruction            & 3.22\% & $-$ & $-$ & $-$  \\
Trigger efficiency              & 0.77\% & $-$ & $-$ & $-$  \\
PID correction algorithm        & 0.20\% & 0.05\%    & 0.06\%    & 0.28\%  \\
Fitting model                   & 1.36\% & 3.67\%    & 2.00\%    & 1.29\% \\
Kinematic reweight              & 0.05\% & $<0.01\%$ & $<0.01\%$ & $<0.01\%$ \\
Statistics of simulated samples & 2.71\% & 4.01\%    & 3.59\%    & 5.58\% \\
 NDC backgrounds                & 1.66\% & 2.44\%    & 0.71\%    & 2.10\% \\
Modeling of \Lc decay amplitude & 1.28\% & 0.09\%    & 1.58\%    & 0.41\% \\
Multiple candidates             & 0.06\% & 1.51\%    & 0.38\%    & 3.44\% \\
\hline
 Total                          & 5.64\% & 6.21\%    & 5.70\%    & 7.35\% \\
\hline
\end{tabular}
}
\end{table}

\section{Significance of signal modes}

To determine the statistical significance of the signal modes, two methods are employed.
For the \Sigmacpp modes, \LbtoScDmK and \LbtoScDstK, the statistical significances are estimated via Wilks' theorem~\cite{Wilks:1938dza}, which relies on the log-likelihood difference, $\Delta\mathcal{L}$, between the default fit and a fit without the \Sigmacpp signal mode.
Under the null (no \Sigmacpp signal) hypothesis, Wilks' theorem specifies that the value of $2\Delta\mathcal{L}$ follows a $\chi^2$ distribution, with a number of degrees-of-freedom equal to the number of additional floating parameters in the default fit.
The $p$ values calculated reject the null hypothesis at significances of $32\sigma$ for \mbox{\LbtoScDmK} and $21\sigma$ for \mbox{\LbtoScDstK}.
Because these are well above the $5\sigma$ observation threshold, the effects of systematic uncertainties are not considered.

For the \Sigmacstarpp modes, pseudoexperiments are used to determine the significances of the \LbtoScstDmK and \LbtoScstDstK signal modes.
The pseudoexperiments are generated from the mass fit model which gives the lowest yield of \mbox{\LbtoScstDmK} (\mbox{\LbtoScstDstK}).
This lowest yield model can be the default model or one of the alternative models, which incorporates systematic uncertainties into the calculation of statistical significance.
The lowest yield model is used to generate 5000 pseudoexperiments with the yields of \Sigmacstarpp signal mode set to zero.
Each pseudoexperiment is then fitted twice, once with the \Sigmacstarpp signal mode yield floating to determine the signal hypothesis log-likelihood, $\mathcal{L}_{\Sigmacstarpp}$, and once without the aforementioned signal component to determine the null hypothesis log-likelihood $\mathcal{L}_{0}$.
The $-2\Delta\mathcal{L} = -2 \times (\mathcal{L}_{0} - \mathcal{L}_{\Sigmacstarpp})$ distribution is then modeled as a $\chi^{2}$ distribution.
Subsequently, the upper tail of this distribution is extrapolated to the data $-2\Delta\mathcal{L}$ value to estimate a $p$ value, which rejects the null hypotheses at significances of $13\sigma$ for \LbtoScstDmK and $9\sigma$ for \LbtoScstDstK.

\section{Conclusion}

The decays \mbox{\LbtoScDmK}, \mbox{\LbtoScDstK}, \mbox{\LbtoScstDmK}, and \mbox{\LbtoScstDstK} have been observed for the first time by employing the LHCb data sample collected during the Run 2 data-taking period, corresponding to an integrated luminosity of 6\invfb. Their measured relative branching fractions are
\begin{align*}
        \frac{\BF\left(\LbtoScDmK\right)}{\BF\left(\LbtoLcDzK\right)}
        & = \RBRresult{0.282}{0.016}{0.016}{0.005},\\
        \frac{\BR(\LbtoScstDmK)}{\BR(\LbtoScDmK)}
        & = \RBRresultNoBrErr{0.460}{0.052}{0.028},\\
        \frac{\BR(\LbtoScDstK)}{\BR(\LbtoScDmK)}
        & = \RBRresult{2.261}{0.202}{0.129}{0.046},\\
        \frac{\BR(\LbtoScstDstK)}{\BR(\LbtoScDmK)}
        & = \RBRresult{0.896}{0.137}{0.066}{0.018},
\end{align*}
where the first uncertainties are statistical, the second are systematic, and the third are due to uncertainties in the branching fractions of intermediate particle decays.
These results provide important inputs for theoretical studies of pentaquark production, in particular in terms of the molecular picture.

These four decay modes only have $\mathcal{O}(100)$ candidates each in the LHCb Run 2 dataset, which is statistically insufficient to perform an amplitude analysis.
This limitation will be overcome with Run 3 data which is expected to increase the statistics by a large factor thanks to the increase in luminosity and trigger efficiency~\cite{LHCb-TDR-012,LHCb-DP-2022-002}.
A future amplitude analysis of these four decay modes will help constrain the characteristics of the three observed pentaquark candidates, which, so far, have only been observed in the discovery channel \mbox{$\Lb\to\jpsi p \Km$}.


\section*{Acknowledgements}
%
%
\noindent We express our gratitude to our colleagues in the CERN
accelerator departments for the excellent performance of the LHC. We
thank the technical and administrative staff at the LHCb
institutes.
We acknowledge support from CERN and from the national agencies:
CAPES, CNPq, FAPERJ and FINEP (Brazil); 
MOST and NSFC (China); 
CNRS/IN2P3 (France); 
BMBF, DFG and MPG (Germany); 
INFN (Italy); 
NWO (Netherlands); 
MNiSW and NCN (Poland); 
MCID/IFA (Romania); 
MICINN (Spain); 
SNSF and SER (Switzerland); 
NASU (Ukraine); 
STFC (United Kingdom); 
DOE NP and NSF (USA).
We acknowledge the computing resources that are provided by CERN, IN2P3
(France), KIT and DESY (Germany), INFN (Italy), SURF (Netherlands),
PIC (Spain), GridPP (United Kingdom), 
CSCS (Switzerland), IFIN-HH (Romania), CBPF (Brazil),
and Polish WLCG (Poland).
We are indebted to the communities behind the multiple open-source
software packages on which we depend.
Individual groups or members have received support from
ARC and ARDC (Australia);
Key Research Program of Frontier Sciences of CAS, CAS PIFI, CAS CCEPP, 
Fundamental Research Funds for the Central Universities, 
and Sci. \& Tech. Program of Guangzhou (China);
Minciencias (Colombia);
EPLANET, Marie Sk\l{}odowska-Curie Actions, ERC and NextGenerationEU (European Union);
A*MIDEX, ANR, IPhU and Labex P2IO, and R\'{e}gion Auvergne-Rh\^{o}ne-Alpes (France);
AvH Foundation (Germany);
ICSC (Italy); 
GVA, XuntaGal, GENCAT, Inditex, InTalent and Prog.~Atracci\'on Talento, CM (Spain);
SRC (Sweden);
the Leverhulme Trust, the Royal Society
 and UKRI (United Kingdom).




\addcontentsline{toc}{section}{References}
\ifx\mcitethebibliography\mciteundefinedmacro
\PackageError{LHCb.bst}{mciteplus.sty has not been loaded}
{This bibstyle requires the use of the mciteplus package.}\fi
\providecommand{\href}[2]{#2}

\newpage
\centerline
{\large\bf LHCb collaboration}
\begin
{flushleft}
\small
R.~Aaij$^{35}$\lhcborcid{0000-0003-0533-1952},
A.S.W.~Abdelmotteleb$^{54}$\lhcborcid{0000-0001-7905-0542},
C.~Abellan~Beteta$^{48}$,
F.~Abudin{\'e}n$^{54}$\lhcborcid{0000-0002-6737-3528},
T.~Ackernley$^{58}$\lhcborcid{0000-0002-5951-3498},
J. A. ~Adams$^{66}$\lhcborcid{0009-0003-9175-689X},
A. A. ~Adefisoye$^{66}$\lhcborcid{0000-0003-2448-1550},
B.~Adeva$^{44}$\lhcborcid{0000-0001-9756-3712},
M.~Adinolfi$^{52}$\lhcborcid{0000-0002-1326-1264},
P.~Adlarson$^{79}$\lhcborcid{0000-0001-6280-3851}
% ...

{\footnotesize \it

% ...
$^{35}$Nikhef National Institute for Subatomic Physics, Amsterdam, Netherlands\\
% ...
$^{44}$Instituto Galego de F{\'\i}sica de Altas Enerx{\'\i}as (IGFAE), Universidade de Santiago de Compostela, Santiago de Compostela, Spain\\
% ...
$^{48}$Physik-Institut, Universit{\"a}t Z{\"u}rich, Z{\"u}rich, Switzerland\\
% ...
$^{52}$H.H. Wills Physics Laboratory, University of Bristol, Bristol, United Kingdom\\
% ...
$^{54}$Department of Physics, University of Warwick, Coventry, United Kingdom\\
% ...
$^{57}$School of Physics and Astronomy, University of Glasgow, Glasgow, United Kingdom\\
$^{58}$Oliver Lodge Laboratory, University of Liverpool, Liverpool, United Kingdom\\
% ...
$^{66}$Syracuse University, Syracuse, NY, United States\\
% ...
$^{79}$Department of Physics and Astronomy, Uppsala University, Uppsala, Sweden, associated to $^{57}$\\
% ...
}
\end{flushleft}

\end{document}